\documentclass[twocolumn, pra, superscriptaddress]{revtex4-1}
\usepackage{graphicx}
\usepackage{dcolumn}
\usepackage{bm}
\usepackage{float}
\usepackage[colorinlistoftodos]{todonotes}
\usepackage{hyperref}
\presetkeys{todonotes}{inline,backgroundcolor=yellow}{}

\usepackage[mathlines]{lineno}

\newcommand{\abcresults}[8]{ #1 & #2 & #5 & #6 & #3 & #4 & #7 & #8 \\}

\modulolinenumbers[5]

\begin{document}

\title{Analysis of Electric Dipole Moment of $^{225}$Ra Atom using the Relativistic Normal Coupled-cluster Theory}

\author{V. S. Prasannaa}
\affiliation{Physical Research Laboratory, Atomic, Molecular and Optical Physics Division, Navrangpura, Ahmedabad-380009, India}

\author{R. Mitra}
\affiliation{Physical Research Laboratory, Atomic, Molecular and Optical Physics Division, Navrangpura, Ahmedabad-380009, India}
\affiliation{Indian Institute of Technology Gandhinagar, Palaj, Gandhinagar-382355, Gujarat, India}

\author{B. K. Sahoo}
\email{bijaya@prl.res.in}
\affiliation{Physical Research Laboratory, Atomic, Molecular and Optical Physics Division, Navrangpura, Ahmedabad-380009, India}

\date{\today}

\begin{abstract} 
    In view of the large differences in the previous calculations of enhancement factors to the parity and time-reversal violating (P,T-odd) electric dipole moment (EDM) of $^{225}$Ra due to nuclear Schiff moment (NSM) and tensor-pseudotensor (T-PT) electron-nucleus (e-N) interactions between the relativistic coupled-cluster (RCC) theory and other many-body methods, we employ the relativistic normal coupled-cluster (RNCC) theory to explain the  discrepancies. The normalization of the wave function in the RNCC theory becomes unity by construction. This feature removes the ambiguity associated with the uncertainties in calculations that could arise due to  mismatch in cancellation of the normalization factor of the wave function in a truncated RCC method. Moreover, all the terms in the expression for EDM using the RNCC method naturally terminate, in contrast to the RCC approach. By taking an average of the results from two variants each of both the RCC and RNCC methods, we recommend enhancement factors to the EDM of $^{225}$Ra due to NSM as $-$6.29(1) $\times 10^{-17}  \arrowvert e \arrowvert$cm $(\arrowvert e \arrowvert fm^3)$ and due to T-PT e-N coupling constant as $-$12.66(14) $\times {10^{-20} \langle \sigma_N \rangle \arrowvert e \arrowvert }$cm, for the nuclear Pauli spinor $\sigma_N$. This is corroborated by analyzing the dipole polarizability ($\alpha_d$) value of $^{225}$Ra, which is obtained as 244(13) $e a_0^3$. We also compare our results for all three properties with previous calculations that employ different many-body methods. Our $\alpha_d$ value agrees very well with the results that are obtained by carrying out rigorous analyses using other variants of RCC methods. 
\end{abstract}

\maketitle

\section{Introduction} \label{sec1}

Atomic systems are excellent candidates to probe electric dipole moments (EDMs), which arise due to parity and time-reversal symmetry (P,T-odd) violating interactions \cite{Krip,dzuba,VVF2}.  Due to the CPT theorem, T violation implies CP violation~\cite{Luders}. Hence, EDMs can also be viewed as arising due to CP violation. The EDMs of diamagnetic atoms, having closed-shell electronic configurations, are more sensitive  to the nuclear Schiff moment (NSM) and the tensor-pseudotensor (T-PT) interactions between the electrons and nucleus (e-N). The CP violating quark-quark (q-q) interactions, as well as the EDMs and chromo-EDMs of quarks, manifest themselves as CP-violating nucleon-nucleon (n-n) interactions and EDMs of nucleons, respectively, which in turn are related to the NSM in atoms~\cite{NSM}. The T-PT e-N interaction, on the other hand, stems from the T-PT electron-quark (e-q) interactions. The Standard Model (SM) of particle physics is known to have withstood many stringent experimental tests till date. However, it is considered as an intermediate manifestation of a complete theory as the SM cannot still explain many observations such as neutrino oscillations, existence of dark matter and dark energy, etc.~\cite{Fukuyama}. It is necessary to have more CP violating interactions than those present in the SM, in order to explain the discrepancy between the observed and the predicted SM value for the matter-antimatter asymmetry in the Universe ~\cite{Sakharov,Can}. The SM also predicts non-zero, but extremely small values for the EDMs, whereas theories beyond the SM predict much larger values for them~\cite{Fukuyama}. Therefore, EDM experiments have become extremely important in probing theories beyond the SM, and also in providing insights into the baryon asymmetry problem. For example, the T-PT e-q interactions will be able to test leptoquark models~\cite{TPT}. 

EDM measurements have been carried out on the $^{129}$Xe, $^{199}$Hg and the radioactive $^{225}$Ra diamagnetic atoms. Recently, two groups have reported improved measurements to the EDM in $^{129}$Xe \cite{expedm,Sachdeva}. They provide improved values over the previous result, which was $(0.7 \pm 3.3_{\text{stat}} \pm 0.1_{\text{sys}}) \times 10^{-27} \arrowvert e \arrowvert$cm with 95\% confidence interval (C.L.) \cite{Rosenberry}. One of the groups obtains the $^{129}$Xe EDM as $(0.26 \pm 2.33_{\text{stat}} \pm 0.07_{\text{sys}}) \times 10^{-27} \arrowvert e \arrowvert$cm (95\% C.L.) \cite{Sachdeva}, and the other one reports it to be $(-4.7 \pm 6.4_{\text{stat}} \pm 0.085_{\text{sys}}$) $\times 10^{-28} \arrowvert e \arrowvert$cm (95\% C.L.) \cite{expedm}, hence improving the earlier result by about three-halves and nine-halves, respectively. In all of the results given above, the subscripts `stat' and `sys' refer to statistical and systematic uncertainties, respectively. However, the most stringent  EDM limit to date comes from $^{199}$Hg as $d_a$($^{199}$Hg) $<$ 7.4 $\times 10^{-30} \arrowvert e \arrowvert$cm (95\% C.L.) \cite{Graner}. The first result for the measurement of EDM in the radioactive $^{225}$Ra atom has been reported with limit $d_a$($^{225}$Ra)$< 1.4 \times 10^{-22}\arrowvert e \arrowvert$cm (95\% C.L.) \cite{expt2}. The advantage of performing an EDM experiment with this atom is that its measurement is enhanced by 3 orders, as compared to $^{129}$Xe and $^{199}$Hg atoms, due to its exceptionally large nuclear octupole deformation \cite{Oct,Spevak}. Thus, the above limit on EDM could be seen as being equivalent to $\sim 10^{-25}\arrowvert e \arrowvert$cm. An improvement of $5$ or $6$ orders in the measurement limit of EDM in $^{225}$Ra could surpass the limit obtained from $^{199}$Hg, and the improved and competitive bound could then be used for inferring the EDMs of quarks and e-q CP violating parameters. Improving this limit is quite possible as $^{225}$Ra has a relatively long lifetime ($\sim 15$ days), and the atom itself can be laser cooled~\cite{Ralc}. Therefore, one could obtain a relatively large number of atoms and a long coherence time for an EDM experiment with $^{225}$Ra in a laser trap, which in turn could improve the statistical sensitivity. Lastly, laser cooled systems also aid in the reduction of the motional magnetic field systematic effect~\cite{expt1}. Nevertheless, it is imperative to measure the EDM in a number of atomic systems, to distinguish contributions from various sources to the EDM, like the NSM, the T-PT e-N interaction, the EDM of the electron, etc.~\cite{Kozlov}. 

It has been demonstrated that in the calculations of enhancement factors in $^{225}$Ra to EDMs due to NSM and T-PT e-N interactions, electron correlation effects are quite large (e.g. see, Ref. \cite{yamanaka,VVF2,dzuba}). Our previous calculations on these quantities by employing a relativistic coupled-cluster (RCC) theory differ by about 40\%, as compared to the results obtained from other variants of relativistic many-body methods \cite{yashpal-Ra}. Though the RCC theory is an all-order perturbative method and it includes electron correlation effects due to core-polarization and other effects on equal footing, the relevant expressions used to evaluate the EDM enhancement factors possess non-terminating terms in the numerator and denominator. If one chooses to directly evaluate these terms, one encounters many disconnected parts that contain the dipole operator ($D$) in the diagrammatic approach, and their inclusions are cumbersome and computationally very expensive. In practical calculations, the numerator is factored into a term containing only connected terms with $D$ and a term with disconnected parts, and the latter is cancelled out with the denominator \cite{yashpal-Hg}. This cancellation is exact only in the complete RCC theory (e.g. see Refs. \cite{cizek,bartlett}). However, this is not the case in an approximated RCC theory, thereby introducing uncertainties in the calculations. This problem can be circumvented by employing the relativistic normal coupled-cluster (RNCC) theory, as demonstrated recently \cite{Sahoo-Hg,Sakurai-Xe}, in which the normalization factor of the wave function does not appear in the calculations of EDM enhancement factors by formulation. Moreover, the expression for determining EDM obeys the Hellman-Feynman theorem and terminates naturally \cite{bishop,bishop1,arponen}. We wish to note at this point that in both Hg as well as Xe, the RCC and the RNCC results were in good agreement with each other\cite{Sahoo-Hg,Sakurai-Xe}. 

In this work, we employ the RNCC theory to evaluate the enhancement factors to EDMs due to NSM and T-PT e-N interactions in $^{225}$Ra. These results are then compared with those from the RCC theory and other many-body methods. We also demonstrate two equivalent, but not really equal, ways of evaluating atomic EDM. We then introduce and elaborate on the RCC and the RNCC theories, and provide the expressions for evaluating a property using these two methods, and by employing the two equivalent approaches in each of them. We also evaluate the static dipole polarizability ($\alpha_d$) of $^{225}$Ra and compare our results with the previously available calculations from various RCC methods.

The manuscript is structured as follows: Sec. \ref{sec2} introduces the reader to the two dominating P and T violating interactions in a closed-shell atom, viz. the T-PT and the NSM, motivates the need for accurate atomic calculations, and their importance in obtaining bounds on properties of interest to new physics that lies beyond the SM. Sec. \ref{sec3} introduces the atomic EDM, and a perturbative approach to treat the P,T-odd interactions of interest to us. Sec. \ref{sec4} discusses the results that we obtain by using these techniques in detail, and compare the results for $\alpha_d$ with previously obtained values in literature that use accurate methods. From this analysis, we give our recommended value for $\alpha_d$. Subsequently, we also explain our approach in obtaining recommended values for $d_a$ due to T-PT as well as NSM. We proceed to expound our analysis of the individual contributions to the final results of $\alpha_d$, $d_a$ due to T-PT, as well as that due to the NSM.

\section{Dominant P,T-odd interactions in $^{225}$Ra} \label{sec2}

The EDM interaction Hamiltonian due to NSM is given by \cite{flambaum,dzuba}
 \begin{eqnarray}
  H_{EDM}^{NSM}= \frac{3{\bf S.r}}{B_4} \rho_N(r),
 \end{eqnarray}
where ${\bf S}=S \frac{{\bf I}}{I}$ is the NSM and $B_4=\int_0^{\infty} dr r^4 \rho_N(r)$. In terms of pion-nucleon-nucleon ($\pi$-n-n) coupling coefficients, $S$ can be expressed as
\cite{haxton, ban, jesus}
\begin{eqnarray}
S = g_{\pi n n} \times (a_0 \bar{g}_{\pi n n}^{(0)} + a_1 \bar{g}_{\pi n n}^{(1)} + a_2 \bar{g}_{\pi n n}^{(2)}),
\end{eqnarray}
where $g_{\pi nn} \simeq 13.5$ is the CP-even $\pi$-n-n coupling constant, $a_i$s are the polarization parameters of the nuclear charge distribution, and $\bar{g}_{\pi n n}^{(i)}$s with $i=$ 1, 2, and 3 represent the isospin components of the CP-odd $\pi$-n-n coupling constants. These couplings are related to the chromo-EDMs of up-quark ($\tilde{d}_u$) and down-quark ($\tilde{d}_d$). Thus, it is necessary to obtain accurate values of $S$ by combining atomic calculations with the experimental EDM result in order to deduce the limits on $\tilde{d}_u$ and $\tilde{d}_d$ from them. 

The T-PT e-N interaction Hamiltonian is given by \cite{yamanaka} 
 \begin{eqnarray}
 H_{EDM}^{T-PT}=i \sqrt{2} G_FC_T \sum_e \bm{\sigma}_N \bm{\cdot \gamma_e} \rho_N(r_e), 
 \end{eqnarray}
where $G_F$ reads as the Fermi constant, $C_T$ is the e-N T-PT coupling constant, {$\bm{\sigma}_N$}$=\langle \sigma_N \rangle \frac{{\bf I}}{I}$ is the Pauli spinor of the nucleus for the nuclear spin $I$, $\rho_N(r)$ is the nuclear density and the subscripts $N$ and $e$ represent the nucleonic and electronic coordinates, respectively. At the elementary particle physics level, $C_T$ can be expressed as a linear combination of electron--up-quark and electron--down-quark T-PT coupling coefficients.  

\section{Perturbed RCC and RNCC Methods}\label{sec3} 

The EDM of the ground state ($\arrowvert \Psi_0 \rangle$) of an atomic system is given by
\begin{eqnarray}
d_a  \equiv \langle D \rangle = \frac{\langle \Psi_0 \arrowvert D \arrowvert \Psi_0 \rangle}{\langle \Psi_0 \arrowvert \Psi_0 \rangle}. 
\end{eqnarray}
Since the permanent EDM arises due to P,T-odd interactions, parity is not a good quantum number to describe atomic states in the presence of these interactions. It is highly non-trivial and challenging to compute such mixed-parity atomic states in the spherical coordinate system. The P,T-odd interactions under consideration are extremely feeble, as compared to the dominant electromagnetic interactions in an atom. Therefore, we can treat a P,T-odd weak interaction as a perturbation over the electromagnetic interactions for evaluating atomic wave functions, by expressing it as 
\begin{equation}
	\arrowvert \Psi_0 \rangle = \arrowvert \Psi_0^{(0)} \rangle + \lambda \arrowvert \Psi_0^{(1)} \rangle^{w} + {\cal O}(\lambda^2),
	\label{eq:psi_0}
\end{equation}
where $\arrowvert \Psi_0^{(0)} \rangle$ is the zeroth-order wave function due to the electromagnetic interactions and $\arrowvert \Psi_0^{(1)} \rangle^{w}$ is the first-order wave function due to a P,T-odd weak interaction. Further, in the above equation, $\lambda$ is the perturbative parameter for the corresponding P,T-odd weak interaction, and the last term ${\cal O}(\lambda^2)$ contains higher-order terms in $\lambda$. The investigated P,T-odd weak interactions are so small that ${\cal O}(\lambda^2)$ contributions could be safely neglected in our analysis. We only consider the Dirac-Coulomb (DC) Hamiltonian ($H^{DC}$) to describe the electromagnetic interactions keeping in mind the required level of accuracy in the present study. 

 After using the above approximation, the expression for $d_a$ yields
 \begin{eqnarray}
\frac{d_a}{\lambda} \simeq 2 \frac{\langle \Psi_0^{(0)} \arrowvert D \arrowvert \Psi_0^{(1)} \rangle^{w}}{\langle \Psi_0^{(0)} \arrowvert \Psi_0^{(0)} \rangle}.
\label{eqpw}
\end{eqnarray}
As mentioned earlier, we intend to demonstrate a credible approach for the numerical evaluation of an atomic EDM. For this purpose, we expand the first-order perturbed wave function in the above expression using a sum-over-states approach; i.e.
\begin{eqnarray}
\frac{d_a}{\lambda} &\simeq& 2 \sum_{I\ne 0} \frac{\langle \Psi_0^{(0)} \arrowvert D \arrowvert \Psi_I^{(0)} \rangle \langle \Psi_I^{(0)} \arrowvert H^{int} \arrowvert \Psi_0^{(0)} \rangle}{\langle \Psi_0^{(0)} \arrowvert \Psi_0^{(0)} \rangle (E_0^{(0)} - E_I^{(0)})} ,
\end{eqnarray}
where $\lambda H^{int}$ is a P,T-odd interaction Hamiltonian, and the summation $I$ is over all possible intermediate states, except the ground state. In the above expression, $E$s with superscripts $(0)$ refer to energies of the unperturbed states. Now, we can rearrange the above expression as 
\begin{eqnarray}
\frac{d_a}{\lambda}&\simeq& 2 \sum_{I\ne 0}  \left [ \langle \Psi_0^{(0)} \arrowvert D \right ] \otimes  \left [ \frac{\arrowvert \Psi_I^{(0)} \rangle \langle \Psi_I^{(0)} \arrowvert H^{int} \arrowvert \Psi_0^{(0)} \rangle}{\langle \Psi_0^{(0)} \arrowvert \Psi_0^{(0)} \rangle (E_0^{(0)} - E_I^{(0)})} \right ] \nonumber \\
&=& 2 \sum_{I\ne 0}  \left [ \frac{\langle \Psi_0^{(0)} \arrowvert D\arrowvert \Psi_I^{(0)} \rangle \langle \Psi_I^{(0)} \arrowvert}{\langle \Psi_0^{(0)} \arrowvert \Psi_0^{(0)} \rangle (E_0^{(0)} - E_I^{(0)})} \right ] \otimes \left [ H^{int} \arrowvert \Psi_0^{(0)} \rangle \right ] \nonumber \\
&=& 2 \sum_{I\ne 0}  \left [ \langle \Psi_0^{(0)} \arrowvert H^{int} \right ] \otimes  \left [ \frac{\arrowvert \Psi_I^{(0)} \rangle \langle \Psi_I^{(0)} \arrowvert D \arrowvert \Psi_0^{(0)} \rangle}{\langle \Psi_0^{(0)} \arrowvert \Psi_0^{(0)} \rangle (E_0^{(0)} - E_I^{(0)})} \right ] \nonumber \\
&=& 2 \frac{\langle \Psi_0^{(0)} \arrowvert H^{int} \arrowvert \Psi_0^{(1)} \rangle^{d}}{\langle \Psi_0^{(0)} \arrowvert \Psi_0^{(0)} \rangle} ,
\label{eqpd}
\end{eqnarray}
where $\arrowvert \Psi_0^{(1)} \rangle^{d}$ represents the first-order perturbed wave function due to $ \lambda D$. It to be noted that we have invoked the hermitian properties of both the operators while rewriting the above expression. Eqs. (\ref{eqpw}) and (\ref{eqpd}) are mathematically equal, but they should be treated as being equivalent from a computational point of view. This is due to the fact that we use an approximated method to determine wave functions and the radial behaviors of both $H^{int}$ and $D$ are very different. Therefore, a comparison of the EDM values obtained from both the approaches by employing a method at a given level of truncation would help in understanding the accuracy of the result and validate the approximation made in that method, by looking for the agreement between the two calculations. It is worth mentioning here that the expression given by Eq. (\ref{eqpw}) should be treated as being more appropriate than Eq. (\ref{eqpd}), while the latter is just a mathematical recast of the former. 

Though we suggest to use both Eqs. (\ref{eqpw}) and (\ref{eqpd}) to test the numerical inaccuracies in the evaluation of $d_a/\lambda$, the difference in our obtained results reflect the imbalance in the calculations of $\arrowvert \Psi_0^{(1)}\rangle^w$ and $\arrowvert \Psi_0^{(1)}\rangle^d$. Therefore, it is also necessary to demonstrate reliability in the determination of $\arrowvert \Psi_0^{(0)} \rangle$ independently. One of the most pertinent approaches to test this is by evaluating a ground state property and comparing the same with its precisely measured experimental result, if available. It can be shown that both Eqs. (\ref{eqpw}) and (\ref{eqpd}) correspond to the expression for $\alpha_d$, when $H^{int}$ is replaced with $D$. Since $D$ is an odd-parity operator and its rank is one, which is the same as that of the considered P,T-odd interaction Hamiltonians, it would be appropriate to calculate $\alpha_d$ and find out its accuracy by using our RCC and RNCC  methods. However, there is no experimental value of $\alpha_d$ that is available for $^{225}$Ra, in order to make a comparison with our calculation. It would still be alright to calculate this property and compare our results with available values that are obtained by rigorous theoretical studies. 

In the RCC theory, the ground state wave function of a closed-shell atom is expressed as
\begin{equation}
	\arrowvert \Psi_0 \rangle = e^{T} \arrowvert \Phi_0 \rangle,
	\label{eqcc}
\end{equation}
where $\arrowvert \Phi_0 \rangle$ is a mean-field wave function, which is obtained by solving the Dirac-Fock (DF) equations in this case, $T$ is RCC operator that takes care of all possible particle-hole excitations arising from $\arrowvert \Phi_0\rangle$. The amplitudes are obtained by solving the following equations 
\begin{eqnarray}
    \langle \Phi_0^* \arrowvert \bar{H}_N \arrowvert \Phi_0 \rangle = 0 ,
\label{ccamp}
\end{eqnarray}
where $H_N=H- \langle \Phi_0 \arrowvert H \arrowvert \Phi_0 \rangle $ is the normal ordered Hamiltonian, a $\arrowvert \Phi_0^*\rangle$ in each equation corresponds to a possible excitation configuration from $\arrowvert \Phi_0\rangle$. In our notation, $\bar{O}=(Oe^T)_c$ for an operator $O$, with the subscript $c$ meaning that the resulting terms in the expression contains connected terms. The energy ($E_0$) of the ground state can be obtained by 
\begin{eqnarray}
   E_0 = \langle \Phi_0 \arrowvert H \arrowvert \Phi_0 \rangle + \langle \Phi_0 \arrowvert \bar{H}_N \arrowvert \Phi_0 \rangle .
   \label{engcc}
\end{eqnarray}

After obtaining $\arrowvert \Psi_0\rangle$, we determine EDM as 
\begin{eqnarray}
    d_a &=& \frac{\langle \Phi_0 \arrowvert e^{T^{\dagger}} D e^T \arrowvert \Phi_0 \rangle}{\langle \Phi_0 \arrowvert e^{T^{\dagger}}  e^T \arrowvert \Phi_0 \rangle} .
\label{prcce}
\end{eqnarray}
Following the linked-diagram theorem, we obtain \cite{cizek,bartlett}
\begin{eqnarray}
 d_a   &=& \frac{{\langle \Phi_0 \arrowvert e^{T^{\dagger}} D_N e^T \arrowvert \Phi_0 \rangle}_c {\langle \Phi_0 \arrowvert e^{T^{\dagger}}  e^T \arrowvert \Phi_0 \rangle}}{\langle \Phi_0 \arrowvert e^{T^{\dagger}}  e^T \arrowvert \Phi_0 \rangle} \nonumber \\
 &=& {\langle \Phi_0 \arrowvert e^{T^{\dagger}} D_N e^T \arrowvert \Phi_0 \rangle}_c .
    \label{prcc}
\end{eqnarray}

However, this is exact only for a complete RCC theory. In an approximated RCC method, cancellation of the normalization factor in Eq. (\ref{prcc}) will not be exact. On the other hand, considering the general expression given by Eq. (\ref{prcce}) demands a large amount of computation because the exponential form of the wave function in the RCC theory could give rise to a large number of disconnected diagrams. Under such circumstances, it is still desirable to use Eq. (\ref{prcc}) for evaluating the EDM in an approximated method as most of the terms associated with these disconnected diagrams in Eq. (\ref{prcce}) almost cancel out leaving behind only fairly small contributions, which could introduce some uncertainties in the calculations, which cannot be precisely predicted a priori, as it also depends on factors such as the choice of system. Nevertheless, the use of either of the Eqs. (\ref{prcce}) and (\ref{prcc}) do not satisfy the Hellmann-Feynman theorem \cite{bishop}. 

The problem discussed above is circumvented in the RNCC method, where the bra state of $\arrowvert \Psi_0 \rangle$ is introduced as \cite{arponen,bishop}
\begin{eqnarray}
 \langle \tilde{\Psi}_0 \arrowvert = \langle \Phi_0 \arrowvert (1+\Lambda) e^{-T} ,
 \label{eqncc}
\end{eqnarray}
where $\Lambda$ is a de-excitation operator like the conjugate $T^{\dagger}$ operator of $T$ in the RCC theory but with different amplitude. It follows that 
\begin{eqnarray}
\langle \tilde{\Psi}_0 \arrowvert \Psi_0 \rangle &=& \langle \Phi_0 \arrowvert (1+\Lambda) e^{-T} e^T \arrowvert \Phi_0 \rangle \nonumber \\
&=& \langle \Phi_0 \arrowvert (1+\Lambda) \arrowvert \Phi_0 \rangle = 1 ,
\label{bio}
\end{eqnarray}
implying that $\langle \tilde{\Psi}_0 \arrowvert$ and $ \arrowvert \Psi_0 \rangle$ satisfy the bi-orthogonal condition. Therefore, it is possible to avoid the appearance of the normalization factor of the wave function while determining any properties, by replacing the bra state of $\arrowvert \Psi_0 \rangle$ with $\langle \tilde{\Psi}_0 \arrowvert$, provided it gives the same energy eigenvalue $E_0$. Now, 
\begin{eqnarray}
 \langle \tilde{\Psi}_0 \arrowvert H \arrowvert \Psi_0 \rangle &=& \langle \Phi_0 \arrowvert (1+\Lambda) e^{-T} H e^T \arrowvert \Phi_0 \rangle  \nonumber \\
 &=&   \langle \Phi_0 \arrowvert H \arrowvert \Phi_0 \rangle + \langle \Phi_0 \arrowvert \bar{H}_N \arrowvert \Phi_0 \rangle \nonumber \\ && + \langle \Phi_0 \arrowvert \Lambda \bar{H} \arrowvert \Phi_0 \rangle \nonumber \\
 &=& E_0 + \langle \Phi_0 \arrowvert \Lambda \bar{H} \arrowvert \Phi_0 \rangle \nonumber \\
  &=& E_0 +\langle \Phi_0 \arrowvert \Lambda \bar{H}_N \arrowvert \Phi_0 \rangle_c.
 \label{engncc} 
\end{eqnarray}
Comparing Eqs. (\ref{engcc}) and (\ref{engncc}), it is obvious that we can get $E_0$ using the RNCC theory by imposing 
 \begin{eqnarray}
 \langle \Phi_0 \arrowvert \Lambda \bar{H}_N \arrowvert \Phi_0 \rangle_c = 0, 
 \label{conncc} 
\end{eqnarray}
besides the bi-orthogonal condition given by Eq. (\ref{bio}). The amplitude equation to be solved for $\Lambda$ is given by 
\begin{eqnarray}
\langle \Phi_0 \arrowvert \Lambda \bar{H}_N \arrowvert \Phi_0^* \rangle &=& - \langle \Phi_0 \arrowvert \bar{H}_N \arrowvert \Phi_0^* \rangle .
\label{nccamp}
\end{eqnarray}
With the knowledge of amplitudes corresponding to the $T$ and $\Lambda$ operators, we can evaluate EDM as 
\begin{eqnarray}
 d_a   &=& \frac{{\langle \Phi_0 \arrowvert \left ( 1+ \Lambda \right ) \bar{D}_N \arrowvert \Phi_0 \rangle}_c} {\langle \Phi_0 \arrowvert \left ( 1 + \Lambda \right ) e^{-T} e^T \arrowvert \Phi_0 \rangle}  \nonumber \\
 &=& {\langle \Phi_0 \arrowvert \Lambda \bar{D}_N \arrowvert \Phi_0 \rangle}_c .
    \label{prncc}
\end{eqnarray}
It is evident that in this expression, the normalization factor of the wave function does not appear by the virtue of its formulation. Eq. (\ref{prncc}) also contains only a finite number of terms. This is also true for an approximated RNCC method. Furthermore, it satisfies the Hellmann-Feynman theorem \cite{bishop}.

\begin{table}[t!]
\caption{\label{tab:table1} The calculated values of the static dipole polarizability, $\alpha_d$ (in units of $ea_0^3$), $d_a$ due to T-PT e-N interaction (in units of $\times 10^{20}C_T \langle \sigma_N \rangle \arrowvert e \arrowvert cm$) and $d_a$ due to NSM (in units of $\times 10^{17} S \arrowvert e \arrowvert cm/(\arrowvert e \arrowvert fm^3)$ using the DF, RCCSD and RNCCSD methods. Methods employed by perturbing the system with weak interaction Hamiltonians and with the dipole operator are denoted by superscripts $w$ and $d$, respectively. Results from previous RCC as well as other many-body theories are also presented for comparison.}
\begin{ruledtabular}
\begin{tabular}{lccc}
 & & \multicolumn{2}{c}{$d_a$ value}\\
 \cline{3-4} \\
 Method & $\alpha_d$ & T-PT&NSM\\ 
 \hline\\
 \multicolumn{4}{l}{\textbf{This work}}\\
 DF        &  204.209 & $-$3.453  &  $-$1.847\\
 RCCSD$^w$ &  230.110 & $-$9.774  & $-$6.183\\
 RCCSD$^d$ &          & $-$12.519 & $-$6.284\\
 RNCCSD$^w$&  256.918 & $-$13.148 & $-$7.053\\
 RNCCSD$^d$&          & $-$12.803 & $-$6.295\\
           &    &          & \\
 Recommended& 244(13)& $-$12.66(14) & $-$6.29(1)\\
 \hline \\
 \multicolumn{4}{l}{\textbf{Previous RCC calculations}}\\
DK+CCSD(T)~\cite{Lim}      & 246.2  &          &          \\
ARPP+CCSD(T)~\cite{Lim2}   & 251.12 &          &          \\
RCCSD(T)~\cite{Anastasia}  & 242.8  &          &          \\
PRCCSD~\cite{Angom}        & 242.42 &          &          \\
RCCSD(T) \cite{yashpal-Ra} & 241.40 & $-$10.01 & $-$6.79  \\
RCCSD \cite{yamanaka}      &        & $-$9.926 & $-$6.215 \\
RCCSD(T)~\cite{BKS1}       & 236    & $-$9     & $-$ 6.1  \\
 \multicolumn{4}{l}{\textbf{Previous RPA calculations}}\\
Ref.~\cite{VVF1}           &        &   $-$17  &  $-$8.3  \\
Ref.~\cite{KVP1}           & 291.4  &  $-$16.59&          \\
Ref.~\cite{VVF2}           & 297    &  $-$8.5  &          \\
Ref.~\cite{yashpal-Ra}     & 296.85 &  $-$16.66& $-$8.12  \\
 \multicolumn{4}{l}{\textbf{Previous CI+MBPT calculations}}\\ 
Ref.~\cite{VVF1}           &        &  $-$18   & $-$8.8    \\
Ref.~\cite{VVF2}           &229.9   &          &           \\
\end{tabular}
\end{ruledtabular}
\end{table}

Having mentioned the differences between the general approaches in the RCC and RNCC theories to determine $d_a$ values, we now turn to the evaluation of the zeroth-order and first-order perturbed wave functions as required by Eqs. (\ref{eqpw}) and (\ref{eqpd}). In the RCC theory, these wave functions can be obtained by expanding the cluster operators as
\begin{eqnarray}
T = T^{(0)} + \lambda T^{(1)} ,
\end{eqnarray}
giving rise to 
\begin{equation}
	\arrowvert \Psi_0^{(0)} \rangle = e^{T^{(0)}} \arrowvert \Phi_0 \rangle,
	\label{cc0}
\end{equation}
and 
\begin{equation}
	\arrowvert \Psi_0^{(1)} \rangle^{w/d} = e^{T^{(0)}} T^{(1)w/d}\arrowvert \Phi_0 \rangle ,
	\label{cc1}
\end{equation}
where the superscripts $w$ and $d$ on the RCC operators stand for perturbation due to the weak interaction Hamiltonian and dipole operator, respectively.

\begin{table}[t]
\caption{\label{tab:table2} Contributions to the static dipole polarizability, $\alpha_d$ (in units of  $ea_0^3$), of $^{225}$Ra from individual terms of the RCCSD and RNCCSD methods. The entry `Others' refers to the terms that are not mentioned in the previous rows. The superscripts $(0)$ and $(1)$ denote the order of perturbation for the corresponding $T$ or $\Lambda$. }
	\begin{tabular}{lc|lc}
	\hline \hline
    RCC term                            & Contribution & RNCC term                            & Contribution \\ \hline\\
	$D {T_1^{(1)}}$                     &135.157   &$D {T_1^{(1)}} $                         &135.157\\
	${T_1^{(1)}}^\dag D$                &135.157   &${\Lambda_1^{(1)}D}$                     &143.200\\
	${T_1^{(0)}}^{\dagger} DT_1^{(1)}$ &$-$10.076 &${\Lambda_1^{(0)w}}^{\dagger} DT_1^{(1)}$&$-$3.286\\ 
	${T_1^{(1)}}^{\dagger} DT_1^{(0)}$ &$-$10.076 &${\Lambda_1^{(1)}}^{\dagger} DT_1^{(0)}$&$-$10.992\\ 
	${T_2^{(0)}}^{\dagger} DT_1^{(1)}$ &$-$15.918 &${\Lambda_2^{(0)}}^{\dagger} DT_1^{(1)}$&0.000\\ 
	${T_1^{(1)}}^{\dagger} DT_2^{(0)}$ &$-$15.918 &${\Lambda_1^{(1)}}^{\dagger} D T_2^{(0)}$&$-$16.856\\
	${T_1^{(0)}}^{\dagger} DT_2^{(1)}$ &1.127     &${\Lambda_1^{(0)}}^{\dagger} DT_2^{(1)}$&0.363\\
	${T_2^{(1)}}^{\dagger} D T_1^{(0)}$&1.127     &${\Lambda_2^{(1)}}^{\dagger} D T_1^{(0)}$&0.000\\
    ${T_2^{(0)}}^{\dagger} DT_2^{(1)}$ &8.008     &${\Lambda_2^{(0)}}^{\dagger} DT_2^{(1)}$&$-$0.145\\
    ${T_2^{(1)}}^{\dagger} DT_2^{(0)}$ &8.008     &${\Lambda_2^{(1)}}^{\dagger} DT_2^{(0)}$&$-$0.227\\
    Others                               &$-$6.486  &Others                                    &9.704\\
    \hline  
    \hline
	\end{tabular}
\end{table}

Substituting $H=H^{DC} + \lambda H^{int}$, or $H=H^{DC} + \lambda D$ and expanding $T$ in Eq. (\ref{ccamp}), we can get the amplitudes of $T^{(0)}$, $T^{(1)w}$ and  $T^{(1)d}$ RCC operators by solving the following equations  
 \begin{eqnarray}
    \langle \Phi_0^* \arrowvert \widetilde{H}^{DC}_N \arrowvert \Phi_0 \rangle &=& 0  , \\
    \langle \Phi_0^* \arrowvert \widetilde{H}^{DC}_N T^{(1)w} \arrowvert \Phi_0 \rangle &=& - \langle \Phi_0^* \arrowvert \widetilde{H}^{int}_N \arrowvert \Phi_0 \rangle  
\label{ccampw}
\end{eqnarray}
and
\begin{eqnarray}
    \langle \Phi_0^* \arrowvert \widetilde{H}^{DC}_N T^{(1)d} \arrowvert \Phi_0 \rangle &=& - \langle \Phi_0^* \arrowvert \widetilde{D}_N \arrowvert \Phi_0 \rangle ,
\label{ccampd}
\end{eqnarray}
respectively, where the notation $\widetilde{O}= (Oe^{T^{(0)}})_c$ is used. After obtaining amplitudes for both the unperturbed and perturbed operators, we can now evaluate the EDM following Eqs. (\ref{eqpw}) and (\ref{prcc}) as
\begin{eqnarray}
 \frac{d_a}{\lambda} &=& 2 {\langle \Phi_0 \arrowvert \overline{D}_N T^{(1)w} \arrowvert \Phi_0 \rangle}_c 
    \label{prccw}
\end{eqnarray}
and following Eqs. (\ref{eqpd}) and (\ref{prcc}) as
\begin{eqnarray}
 \frac{d_a}{\lambda} &=& 2 {\langle \Phi_0 \arrowvert \overline{H}^{int}_N T^{(1)d} \arrowvert \Phi_0 \rangle}_c 
    \label{prccd}
\end{eqnarray}
where $\overline{O}= e^{T^{(0)\dagger}} Oe^{T^{(0)}}$. As mentioned earlier, both Eqs. (\ref{prccw}) and (\ref{prccd}) are the same but a careful inspection shows that the former involves a non-terminating series in $\overline{D}_N$ and $T^{(1)w}$ with finite terms, while the latter also involves an infinite series, but in $\overline{H}^{int}_N$ and $T^{(1)d}$ contains finite terms. This means that $H^{int}$ and also $D$ are not treated equally in both Eqs. (\ref{prccw}) and (\ref{prccd}), which may cause differences in the results when an approximated RCC theory is employed. 

\begin{figure}[t]
\centering
\includegraphics[width=8.5cm, height=8.5cm]{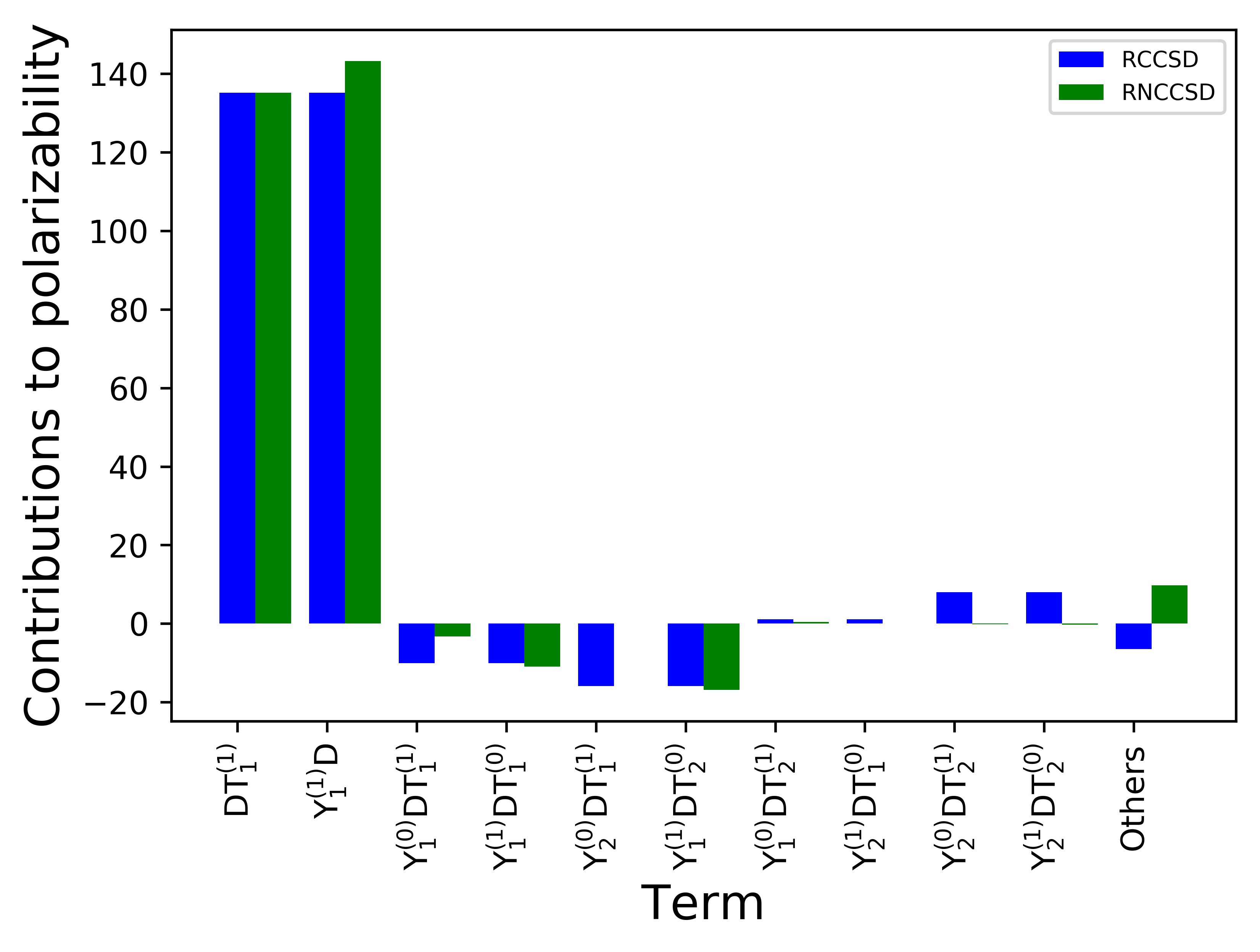} 
\caption{A schematic representation of the contributions to $\alpha_d$ in $^{225}$Ra from RCCSD and RNCCSD methods. Here, $Y$ is used to denote the $T^{\dagger}$ operator of the RCCSD method and the $\Lambda$ operator of the RNCCSD method. The superscripts $(0)$ and $(1)$ denote the order of perturbation for the corresponding $Y$ or $T$. } 
\label{fig1}
\end{figure}

\begin{table*}[t]
\caption{\label{tab:table3} Comparison of contributions to the $d_a$ value (in units of $\times 10^{20}C_T \langle \sigma_N \rangle \arrowvert e \arrowvert cm$) of $^{225}$Ra due to T-PT e-N interaction among individual terms of the RCCSD and RNCCSD methods. `Others' refers to the terms that are not mentioned here. The notation with the superscripts $(0)$ and $(1)$ is the same as in the previous table. The results obtained by perturbing the system with the weak P,T-odd $H^{int}$ (in this case, the T-PT Hamiltonian) is denoted by the superscript $w$, while the superscript $d$ is used when the dipole operator $D$ is treated as a perturbation on the RCC and RNCC operators. } 
	\begin{tabular}{lclc|lclc}
	\hline \hline
	\multicolumn{4}{c|}{Perturbing by P,T-odd Hamiltonian}&  \multicolumn{4}{c}{Perturbing by dipole operator} \\
\cline{1-4}\cline{5-8}\\
	 RCC term & Contribution & RNCC term & Contribution & RCC term & Contribution & RNCC term & Contribution \\
	\hline \\
	\abcresults{$D{T_1^{(1)w}}$}{$-$6.467}{$H^{int}{T_1^{(1)d}}$}{$-$5.427}{$D{T_1^{(1)w}}$}{$-$6.467}{$H^{int} {T_1^{(1)d}}$}{$-$5.427}
	\abcresults{${T_1^{(1)w}}^\dag D$}{$-$6.467}{${T_1^{(1)d}}^\dag H^{int}$}{$-$5.427}{${\Lambda_1^{(1)w}D}  $}{$-$8.517}{${\Lambda_1^{(1)d}H^{int}}$}{$-$5.754}
	\abcresults{${T_1^{(0)}}^{\dagger}DT_1^{(1)w}$}{$-$0.055}{${T_1^{(0)}}^{\dagger} H^{int}T_1^{(1)d}$}{$-$0.643}{${\Lambda_1^{(0)}}^{\dagger}DT_1^{(1)w}$}{$-$0.022}{$ {\Lambda_1^{(0)}}^{\dagger}H^{int}T_1^{(1)d}$}{$-$0.277} 
	\abcresults{${T_1^{(1)w}}^{\dagger}DT_1^{(0)w}$}{$-$0.055}{${T_1^{(1)d}}^{\dagger} H^{int}T_1^{(0)}  $}{$-$0.643}{${\Lambda_1^{(1)w}}^{\dagger}DT_1^{(0)}$}{0.013}{$ {\Lambda_1^{(1)d}}^{\dagger} H^{int}T_1^{(0)}$}{$-$0.735}
	\abcresults{${T_2^{(0)}}^{\dagger} DT_1^{(1)w} $}{1.656}{${T_2^{(0)}}^{\dagger} H^{int}T_1^{(1)d}$}{$-$0.972}{${\Lambda_2^{(0)}}^{\dagger} DT_1^{(1)w}$}{0.000}{$ {\Lambda_2^{(0)}}^{\dagger} H^{int}T_1^{(1)d} $}{0.000} 
	\abcresults{${T_1^{(1)w}}^{\dagger} DT_2^{(0)} $}{1.656}{$ {T_1^{(1)d}}^{\dagger} H^{int} T_2^{(0)} $}{$-$0.972}{$ {\Lambda_1^{(1)w}}^{\dagger} D T_2^{(0)} $}{1.962}{$ {\Lambda_1^{(1)d}}^{\dagger} H^{int} T_2^{(0)} $}{$-$1.033}
	\abcresults{${T_1^{(0)}}^{\dagger} DT_2^{(1)w} $}{$-$0.044}{$ {T_1^{(0)}}^{\dagger} H^{int}T_2^{(1)d} $}{$-$0.046}{$ {\Lambda_1^{(0)}}^{\dagger} DT_2^{(1)w} $}{$-$0.018}{$ {\Lambda_1^{(0)}}^{\dagger} H^{int}T_2^{(1)d} $}{$-$0.019}
	\abcresults{${T_2^{(1)w}}^{\dagger} D T_1^{(0)} $}{$-$0.044}{$ {T_2^{(1)d}}^{\dagger} H^{int} T_1^{(0)}$}{$-$0.046}{$ {\Lambda_2^{(1)w}}^{\dagger} D T_1^{(0)} $}{0.000}{$ {\Lambda_2^{(1)d}}^{\dagger} H^{int} T_1^{(0)}$}{0.000}
    \abcresults{${T_2^{(0)}}^{\dagger} DT_2^{(1)w} $}{$-$0.078}{${T_2^{(0)}}^{\dagger} H^{int}T_2^{(1)d} $}{$-$0.035}{${\Lambda_2^{(0)}}^{\dagger} DT_2^{(1)w} $}{$-$0.014}{${\Lambda_2^{(0)}}^{\dagger} H^{int}T_2^{(1)d}$}{$-$0.037}
    \abcresults{${T_2^{(1)w}}^{\dagger} DT_2^{(0)} $}{$-$0.078}{${T_2^{(1)d}}^{\dagger} H^{int}T_2^{(0)} $}{$-$0.035}{${\Lambda_2^{(1)w}}^{\dagger} DT_2^{(0)} $}{0.047}{${\Lambda_2^{(1)d}}^{\dagger} H^{int}T_2^{(0)} $}{$-$0.064}
    \abcresults{Others}{0.202}{Others}{1.726}{Others}{$-$0.131}{Others}{0.543}
    \hline
    \hline
	\end{tabular}
\end{table*}

We now turn to a perturbative approach to evaluate the atomic EDM using the RNCC theory. Pursuing similarly as in the RCC theory, we can express  
\begin{eqnarray}
\Lambda = \Lambda^{(0)} + \lambda \Lambda^{(1)w} 
\end{eqnarray}
and 
\begin{eqnarray}
\Lambda = \Lambda^{(0)} + \lambda \Lambda^{(1)d} 
\end{eqnarray}
by considering $\lambda H^{int}$ and $D$ as the perturbation, respectively. Following Eq. (\ref{nccamp}), we can get the amplitude equations for the $\Lambda^{(0)}$, $\Lambda^{(1)w}$ and  $\Lambda^{(1)d}$ RNCC operators as 
 \begin{eqnarray}
    \langle \Phi_0 \arrowvert \Lambda^{(0)} \widetilde{H}^{DC}_N \arrowvert \Phi_0^* \rangle &=& - \langle \Phi_0 \arrowvert \widetilde{H}^{DC}_N \arrowvert \Phi_0^* \rangle , \\
    \langle \Phi_0 \arrowvert \Lambda^{(1)w} \widetilde{H}^{DC}_N \arrowvert \Phi_0^* \rangle &=& - \langle \Phi_0 \arrowvert \widetilde{H}^{int}_N \left ( 1 + \Lambda^{(0)}\right ) \nonumber \\ && + \widetilde{H}^{DC}_N T^{(1)w}\arrowvert \Phi_0^* \rangle  
\label{nccampw}
\end{eqnarray}
and
\begin{eqnarray}
 \langle \Phi_0 \arrowvert \Lambda^{(1)d} \widetilde{H}^{DC}_N \arrowvert \Phi_0^* \rangle &=& - \langle \Phi_0 \arrowvert \widetilde{D}_N \left ( 1 + \Lambda^{(0)}\right ) \nonumber \\ && + \widetilde{H}^{DC}_N T^{(1)d}\arrowvert \Phi_0^* \rangle, 
\label{nccampd}
\end{eqnarray}
respectively. It follows from Eq. (\ref{conncc}) that the conditions 
 \begin{eqnarray}
 \langle \Phi_0 \arrowvert \Lambda^{(1)w} \bar{H}^{DC}_N + \left ( 1+ \Lambda^{(0)} \right ) \bar{H}^{int}_N \arrowvert \Phi_0 \rangle_c = 0
 \label{conncc1} 
\end{eqnarray}
and 
\begin{eqnarray}
  \langle \Phi_0 \arrowvert \Lambda^{(1)d} \bar{H}^{DC}_N + \left ( 1+ \Lambda^{(0)} \right ) \bar{D}_N \arrowvert \Phi_0 \rangle_c = 0
 \label{conncc2} 
\end{eqnarray}
must be imposed, while solving for the amplitudes of $\Lambda^{(1)w}$ and $\Lambda^{(1)d}$, respectively. This yields the expressions for evaluating the EDM after perturbing with $H^{int}$ and $D$ as
\begin{eqnarray}
 \frac{d_a}{\lambda} &=& {\langle \Phi_0 \arrowvert \Lambda^{(1)w} \widetilde{D}_N + \left (1+ \Lambda^{(0)} \right ) \widetilde{D}_N T^{(1)w} \arrowvert \Phi_0 \rangle}_c 
    \label{prnccw}
\end{eqnarray}
and
\begin{eqnarray}
 \frac{d_a}{\lambda} &=& {\langle \Phi_0 \arrowvert \Lambda^{(1)d} \widetilde{H}^{int}_N + \left (1+ \Lambda^{(0)} \right ) \widetilde{H}^{int}_N T^{(1)d} \arrowvert \Phi_0 \rangle}_c ,
    \label{prnccd}
\end{eqnarray}
respectively. In the above expressions, the factor $2$ does not appear due to the fact that both the ket and bra states are treated differently. Here, both Eqs. (\ref{prnccw}) and (\ref{prnccd}) contain naturally terminating expressions and interchange of $D$ and $H^{int}$ should, in principle, give identical results. However, it is possible to encounter large uncertainties while treating $H^{int}$ or $D$ as a perturbation in order to obtain the first-order correction to the wave function due to their different convergence behaviors. 

Again, we can determine $\alpha_d$ values by replacing $H^{int}$ by $D$ in both the perturbed RCC and RNCC theories. As a result, the asymmetry between $H^{int}$ and $D$ appearing in these theories cannot be distinguished for this property. We consider all possible one particle-one hole (denoted by subscript $1$) and two particle-two hole (denoted by subscript $2$) excitations in both the RCC theory (RCCSD method) and the RNCC theory (RNCCSD method), in which case the excitation operators are given by 
\begin{eqnarray}
&& T^{(0/1)} = T_1^{(0/1)} + T_2^{(0/1)} \nonumber \\ \text{and} && \nonumber \\ && \Lambda^{(0/1)} = \Lambda_1^{(0/1)} + \Lambda_2^{(0/1)}.
\end{eqnarray}

For the calculations of the EDM enhancement factors due to NSM and T-PT e-N interaction,
we use $\lambda= 1.0 \times 10^{-17} S \arrowvert e \arrowvert cm/(\arrowvert e \arrowvert fm^3)$ and $\lambda=1.0 \times 10^{-20} C_T \langle \sigma_N \rangle \arrowvert e \arrowvert cm$, respectively, while we choose $\lambda=1$ for the evaluation of $\alpha_d$. We present results from both the RCCSD and RNCCSD methods in the following section. 

\section{Results and discussion} \label{sec4}

We use Gaussian type orbitals (GTOs) \cite{gtos1,gtos2,gtos3} to define the large and small components of the single-particle orbitals to carry out our calculations. These GTOs are optimized by comparing bound orbitals with those obtained numerically by employing the General Relativistic Atomic Structure Program (GRASP2) \cite{grasp}. All together, we produce orbitals up to a principal quantum number of $n=40$ up to $l=4$ angular momentum symmetry. Among these, excitations of electrons among 20 s-orbitals, 19 p-orbitals, 18 d-orbitals, 16 f-orbitals and 12 g-orbitals are carried out both in the RCCSD and RNCCSD methods. Contributions from beyond these orbitals are investigated using a second-order perturbation theory and they are found to be negligibly small. It is worth adding here that the energy cut-off values were as high as about 3500 atomic units for s, 2500 for p, 1800 for d, 600 for f, and 500 for the g symmetries. Such careful choice of values could ensure that the basis functions in each symmetry are spread out evenly across a wide energy range. 

In Table \ref{tab:table1}, we present $\alpha_d$ and $d_a$ values due to the T-PT e-N interaction and NSM that we obtained from both the RCCSD and RNCCSD methods. To distinguish the results obtained by perturbing the system with the weak T-PT or NSM interaction from those obtained by perturbing with the dipole interaction, we denote the methods that use the former with the superscript $w$ and those with the latter by using the superscript $d$. As mentioned in Sec. \ref{sec1}, we report our results by using the RCCSD$^w$ method and compare them with results obtained using other many-body methods in Ref. \cite{yashpal-Ra}. In Table \ref{tab:table1}, we compare the results with earlier RCC calculations. For completeness, we have also added previous works that employ other many-body methods, such as random phase approximation (RPA)~\cite{VVF1,KVP1,VVF2,yashpal-Ra} and a hybrid approach of configuration interaction method with many-body perturbation theory (CI+MBPT)~\cite{VVF1,VVF2} to the above table. The RPA results, calculated using a $V^N$ potential, are subsumed in the RCC framework. The CI+MBPT hybrid approach employs a $V^{(N-2)}$ potential, while ours uses a $V^N$ potential to obtain the single-particle orbitals. Moreover, their approach does not capture electron correlations as effectively as a RCC approach, which becomes crucial for Ra due to its pronounced correlation effects. Turning our attention to the earlier RCC calculations, we find that the they are very similar to our present results; with the minor differences attributed to the use of more optimized GTOs in this work. Nonetheless, it can be noticed that the $d_a$ values due to both the T-PT and NSM interactions obtained using the RCCSD$^w$ and RCCSD$^d$ approaches differ substantially. As stated in Sec. \ref{sec3}, in these approaches, $H^{int}$ and $D$ are treated differently; and hence their radial behaviors are completely dissimilar. These can be the reasons for the discrepancies between the above two procedures in the RCCSD method. In contrast, we find less discrepancies among the $d_a$ values due to T-PT that are obtained by adopting the RNCCSD$^w$ and RNCCSD$^d$ approaches. This could be due to the fact that RNCCSD method does not involve any non-terminating expressions. However, the difference between the $d_a$ value due to NSM between the RNCCSD$^w$ and RNCCSD$^d$ approaches are found to be significant. On other hand, $d_a$ values between the RCCSD$^d$ and RNCCSD$^d$ approaches are seen to be agreeing quite well while large differences are observed between the results from the RCCSD$^w$ and RNCCSD$^w$ approaches.

\begin{figure}[t]
\centering
\includegraphics[width=8.5cm, height=8.5cm]{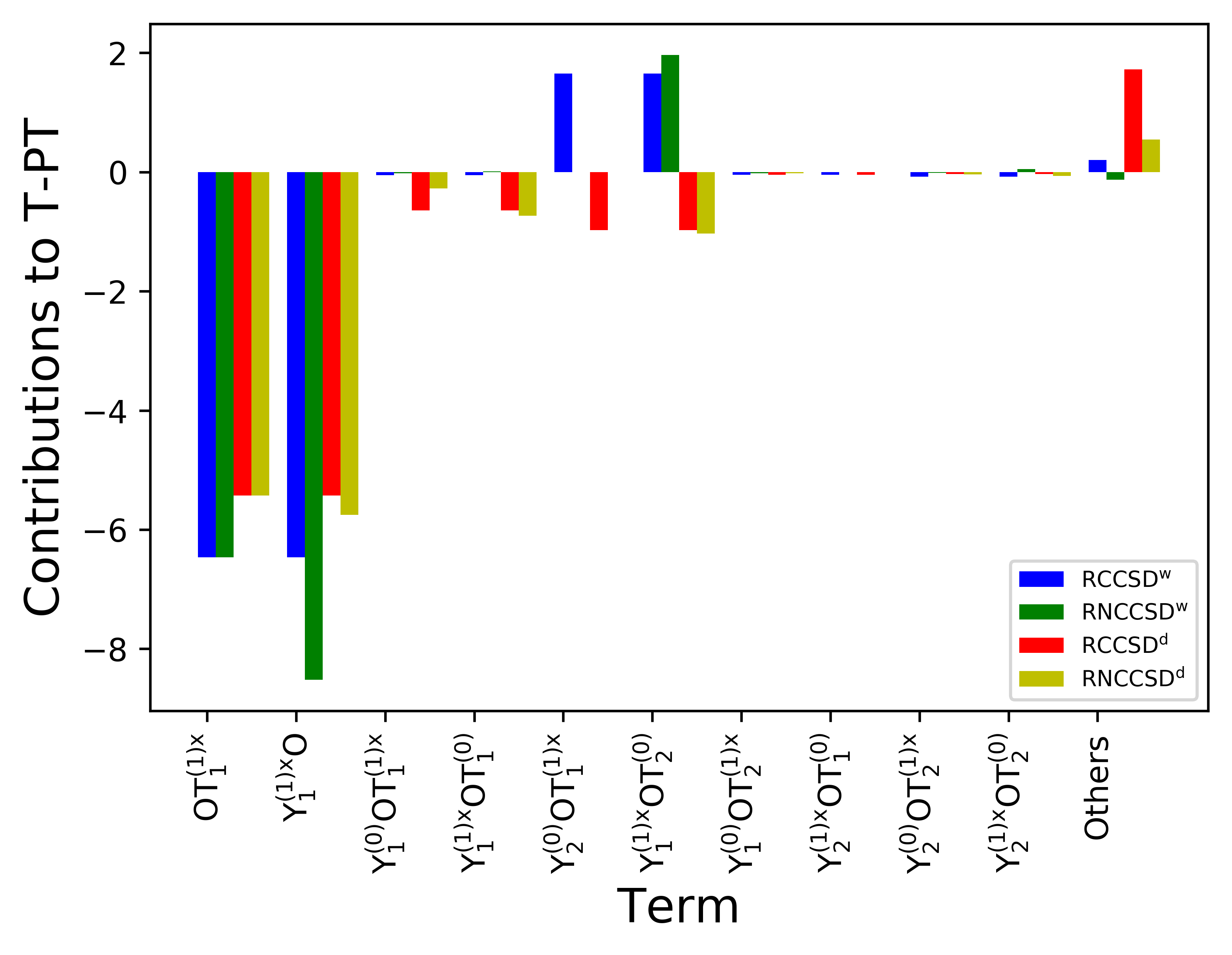}
\caption{A pictorial illustration of contributions from various RCCSD and RNCCSD methods to $d_a$ due to T-PT e-N interaction in $^{225}$Ra. Values are scaled by ($10^{20}C_T \langle \sigma_N \rangle \arrowvert e \arrowvert cm$). Here, $Y$ is used as a dummy operator, and is the $T^{\dagger}$ operator in the RCCSD method or the $\Lambda$ operator in the RNCCSD method. The notations for the superscripts $(0)$ and $(1)$ are the same as in the previous figure. The superscript $x$ is $w$ when we perturb with the weak interaction Hamiltonians, and is $d$ in the case of using the dipole operator. The operator $O$ is either the T-PT operator or the dipole operator, depending on whether $x$ is $d$ or $w$, respectively. }
\label{fig2}
\end{figure}

In Sec. \ref{sec3}, we had mentioned that the results obtained by the RCCSD$^w$ and RNCCSD$^w$ approaches are to be regarded as being the correct way of determining $d_a$ values in atomic systems. However, the above finding demonstrates that these results may not be reliable. As mentioned earlier, it would be appropriate to determine a physical property using the above methods and compare the results with its experimental value to understand the reliability of the calculations. From this point of view, the calculation of $\alpha_d$ would be very appropriate to make such a comparison. We have obtained the value of $\alpha_d$ as 230.110 $ea_0^3$ and 256.918 $ea_0^3$ from the RCCSD and RNCCSD methods, and they differ substantially. Compared to the DF value of 204.209 $ea_0^3$, the RCCSD result seems to only add a relatively small correlation contribution while in the RNCCSD method, the contribution is substantial. It is worth noting here that we have adopted Exclusion-Principle-Violating (EPV) diagrams in the implementation of the RCCSD and RNCCSD methods \cite{lindgrenbook}. In such a case, unphysical contributions cancel out through the direct and exchange terms due to the Fermi-Dirac statistics. Since the employed RCCSD method involves non-terminating terms in the determination of $\alpha_d$ and $d_a$ values, it can introduce unphysical contributions in the calculations due to brute-force termination of the expression. Unfortunately, there is no experimental value of $\alpha_d$ available for Ra atom to test the validity of our RCCSD and RNCCSD results. Therefore, it would be necessary to compare our calculations with other theoretical studies, mainly in the framework of RCC theory, which have analyzed electron correlation effects in $\alpha_d$ rigorously. 

\begin{table*}[t]
\caption{\label{tab:table4} Comparison of contributions the $d_a$ value (in $10^{17} S \arrowvert e \arrowvert cm/(\arrowvert e \arrowvert fm^3)$  of $^{225}$Ra due to NSM interaction between the individual terms of the RCCSD and RNCCSD methods. The notation in use is borrowed from the previous table, with $H^{int}$ being the NSM interaction Hamiltonian here. } 
	\begin{tabular}{lclc|lclc}
	\hline \hline
	\multicolumn{4}{c|}{Perturbing by P,T-odd Hamiltonian}&  \multicolumn{4}{c}{Perturbing by dipole operator} \\
\cline{1-4}\cline{5-8}\\
	RCC term & Contribution & RNCC term & Contribution & RCC term & Contribution & RNCC term & Contribution \\
\hline \\	
	\abcresults{$D{T_1^{(1)w}}$}{$-$3.770}{$H^{int}{T_1^{(1)d}}$}{$-$2.720}{$D{T_1^{(1)w}}$}{$-$3.770}{$H^{int} {T_1^{(1)d}}$}{$-$2.720}
	\abcresults{${T_1^{(1)w}}^\dag D$}{$-$3.770}{${T_1^{(1)d}}^\dag H^{int}$}{$-$2.720}{${\Lambda_1^{(1)w}D}  $}{$-$3.931}{${\Lambda_1^{(1)d}H^{int}}$}{$-$2.705}
	\abcresults{${T_1^{(0)w}}^{\dagger}DT_1^{(1)w}$}{0.003}{${T_1^{(0)d}}^{\dagger} H^{int}T_1^{(1)d}$}{$-$0.315}{${\Lambda_1^{(0)w}}^{\dagger}DT_1^{(1)w}$}{0.001}{$ {\Lambda_1^{(0)d}}^{\dagger}H^{int}T_1^{(1)d}$}{$-$0.155} 
	\abcresults{${T_1^{(1)w}}^{\dagger}DT_1^{(0)w}$}{0.003}{${T_1^{(1)d}}^{\dagger} H^{int}T_1^{(0)d}$}{$-$0.315}{${\Lambda_1^{(1)w}}^{\dagger}DT_1^{(0)w}$}{0.015}{$ {\Lambda_1^{(1)d}}^{\dagger} H^{int}T_1^{(0)d}$}{$-$0.337} 
	\abcresults{${T_2^{(0)w}}^{\dagger} DT_1^{(1)w} $}{0.768}{${T_2^{(0)d}}^{\dagger} H^{int}T_1^{(1)d}$}{$-$0.519}{${\Lambda_2^{(0)w}}^{\dagger} DT_1^{(1)w}$}{0.000}{$ {\Lambda_2^{(0)d}}^{\dagger} H^{int}T_1^{(1)d} $}{0.000} 
	\abcresults{${T_1^{(1)w}}^{\dagger} DT_2^{(0)w} $}{0.768}{$ {T_1^{(1)d}}^{\dagger} H^{int} T_2^{(0)d}$}{$-$0.519}{${\Lambda_1^{(1)w}}^{\dagger} D T_2^{(0)w}$}{0.796}{$ {\Lambda_1^{(1)d}}^{\dagger} H^{int} T_2^{(0)d}$}{$-$0.561}
	\abcresults{${T_1^{(0)w}}^{\dagger} DT_2^{(1)w} $}{0.023}{$ {T_1^{(0)d}}^{\dagger} H^{int}T_2^{(1)d} $}{$-$0.026}{$ {\Lambda_1^{(0)w}}^{\dagger} DT_2^{(1)w} $}{0.008}{$ {\Lambda_1^{(0)d}}^{\dagger} H^{int}T_2^{(1)d} $}{$-$0.012}
	\abcresults{${T_2^{(1)w}}^{\dagger} D T_1^{(0)w} $}{0.023}{$ {T_2^{(1)d}}^{\dagger} H^{int} T_1^{(0)d} $}{$-$0.026}{${\Lambda_2^{(1)w}}^{\dagger} D T_1^{(0)w}$}{0.000}{$ {\Lambda_2^{(1)d}}^{\dagger} H^{int} T_1^{(0)d}$}{0.000}
    \abcresults{${T_2^{(0)w}}^{\dagger} DT_2^{(1)w} $}{$-$0.207}{${T_2^{(0)d}}^{\dagger} H^{int}T_2^{(1)d}$}{$-$0.002}{${\Lambda_2^{(0)w}}^{\dagger} DT_2^{(1)w} $}{0.004}{${\Lambda_2^{(0)d}}^{\dagger} H^{int}T_2^{(1)d}$}{$-$0.039}
    \abcresults{${T_2^{(1)w}}^{\dagger} DT_2^{(0)w} $}{$-$0.207}{${T_2^{(1)d}}^{\dagger} H^{int}T_2^{(0)d}$}{$-$0.002}{${\Lambda_2^{(1)w}}^{\dagger} DT_2^{(0)w} $}{0.013}{${\Lambda_2^{(1)d}}^{\dagger} H^{int}T_2^{(0)d}$}{$-$0.055}
    \abcresults{Others}{0.182}{Others}{0.880}{Others}{$-$0.188}{Others}{0.290}
    \hline  \hline
	\end{tabular}
\end{table*}

As can be seen from Table \ref{tab:table1}, our RCCSD result for $\alpha_d$ agrees with the value obtained using CI+MBPT method. However, the RNCCSD result seems to be closer to the other RCC calculations as discussed below. Lim \textit{et al}~\cite{Lim} computed $\alpha_d$ by using the Douglas-Kroll Hamiltonian and incorporating spin-orbit coupling interaction. They had employed contracted GTOs and reported an $\alpha_d$ of 248.56 $ea_0^3$ using the coupled-cluster with singles, doubles and partial triples (CCSD(T) method) approximation. Their final recommended value was (246.2 $\pm$ 4.9) $ea_0^3$. Later, they had revised its value to 251.12 $ea_0^3$ \cite{Lim2} by using a j-averaged scalar relativistic pseudo potential (ARPP) and allowing correlation among the  outer-core and valence electrons in the CCSD(T) method. Borschevsky \textit{et al}~\cite{Anastasia} performed calculations to obtain $\alpha_d$ using an uncontracted universal basis set in the relativistic CCSD(T) method (RCCSD(T) method). They also employed the DC Hamiltonian and the Gaunt term, to obtain a final result of 242.8 $ea_0^3$. All these RCC methods adopt the finite-field approach to evaluate $\alpha_d$. Another calculation by Chattopadhyay \textit{et al}~\cite{Angom} reported its value as 242.42 $ea_0^3$ using the perturbed RCC theory like ours (denoted by PRCC method) by considering the Dirac-Coulomb-Breit Hamiltonian. Our earlier calculation using RCCSD method and including the corrections from  partial triples give its value as 241.40 $ea_0^3$ \cite{yashpal-Ra}. 

All the above theoretical calculations show that $\alpha_d$ value would lie in between 240 and 250 $ea_0^3$. By taking an average of the values from the RCCSD and RNCCSD methods employed in this work, we get an $\alpha_d$ of 243.514 $ea_0^3$. Taking into account deviation of RCCSD and RNCCSD results from the average value, we recommend a value of 244(13) $ea_0^3$ for $\alpha_d$ of the Ra atom. This value is in agreement with the above findings. From these analyses of the $\alpha_d$ result, we draw the conclusion that RCCSD method gives a lower estimate to a possible true value. It may be due to unequal cancellation of unphysical contributions arising through EPV diagrams due to forcibly terminating an otherwise infinite series appearing in this method. On the other hand, the RNCCSD approach overestimates $\alpha_d$. The inclusion of triple excitations in both the RCC and RNCC theories may lead to a much better agreement. We would like to emphasize on the fact that trends in the contributions from triple excitations through RCC theory will be very different than those from the RNCC theory. However, implementations of full triple excitations in these methods by adopting a spherical coordinate system are very challenging because of involvement of complicated angular momentum couplings. On the basis of these considerations, we would offer recommended values of $d_a$ due to the T-PT e-N interaction and NSM by taking the average of the results from the RCCSD and RNCCSD methods. Since $d_a$ values differ substantially when they are evaluated by considering the weak interaction Hamiltonian as a perturbation to the wave function and the case where the dipole operator is chosen instead, it would be necessary to understand the possible origins of errors that cause these discrepancies. One such source, as already mentioned earlier, is the forcible termination of the expression appearing in the RCC theory. Errors could stem from the large numerical uncertainties associated with equations that determine the amplitudes of the first-order perturbed wave functions. We have adopted a  Jacobi-iterative scheme to solve Eqs. (\ref{ccampw}), (\ref{ccampd}), (\ref{nccampw}) and (\ref{nccampd}). We observe that for the RCC$^w$ case, the amplitudes turn out to be unusually large in the first few  iterations, especially for the NSM interaction, and they gradually reduce and converge to a very small value after a sufficiently long period of time. On the other hand, the first-order perturbed wave functions due to dipole operator converge smoothly. This may be the main reason for observing consistent $d_a$ results from the RCCSD$^d$ and RNCCSD$^d$ approaches. On this basis, we consider these results to be more reliable. Now, proceeding in a similar manner as in the case of the  $\alpha_d$ result, we recommend $d_a$ values as $-12.66(14) \times 10^{-20} C_T \langle \sigma_N \rangle \arrowvert e \arrowvert cm$ and $-6.29(1) \times 10^{-17}S \arrowvert e \arrowvert cm/(\arrowvert e \arrowvert fm^3)$ due to T-PT e-N and NSM interactions, respectively. Here, we have neglected the contributions to $d_a$ values due to Breit and quantum electrodynamics (QED) interactions as previous calculations suggest that they add little to the results obtained without these interactions \cite{yashpal-Ra}. 

We now turn to the trends in electron correlation effects from different approximations in order to understand the possible reasons of discrepancies in the results and gain insights into various contributions to the investigated properties. In Table \ref{tab:table2}, we present the  individual contributions to $\alpha_d$ from the RCCSD and the RNCCSD methods. Here, we drop superscripts $w$ and $d$ on the RCC and RNCC operators as they do not serve any purpose. We use a shorthand notation both in the table and in the text that follows, for ease of description. It is immediately evident from the table that the dominant contributions are coming from $D {T_1^{(1)}}$, and its hermitian conjugate (h.c.) for the RCCSD method or $\Lambda_1^{(1)}D$ for the RNCCSD method. The contribution from $\Lambda_1^{(1)}D$ is found to be larger than $D {T_1^{(1)}}$. This trend is opposite to that found in $^{129}$Xe and $^{199}$Hg, as shown in recent studies \cite{Sakurai-pol,Sahoo-Hg}. These terms contain the lowest-order DF contributions, therefore the differences between the DF results and contributions from these terms are due to the correlation effects. As described in the previous studies \cite{yashpal-polz,Sakurai-pol}, the leading order pair-correlation and core-polarization effects are incorporated through these terms.  The next dominant set of terms are $T_{1/2}^{(1/0)}DT_{2/1}^{(0/1)}$ in the RCCSD method, and their counterparts in the RNCCSD method. The slash denotes that the first and the second subscripts are either $1$ or $2$, respectively, or vice-versa, and the corresponding first and the second superscripts are either $(1)$ or $(0)$, respectively, or vice-versa. These are the higher-order correlation contributions. On further examination of the table, we see the interplay between the different terms and their signs that in turn explain why the result from the RCCSD method is much lesser than that obtained using the RNCCSD approach. The terms $T_{1}^{(0)}DT_{1}^{(1)}$ through $T_{1}^{(1)}DT_{2}^{(0)}$ in the RCCSD method add to about $-$52 $ea_0^3$, while the terms $\Lambda_{1}^{(0)}DT_{1}^{(1)}$ through $\Lambda_{1}^{(1)}DT_{2}^{(0)}$ in the RNCCSD method add to about $-$31 $ea_0^3$ to their final values of $\alpha_d$. This is attributed to the rather large `asymmetry' between the contributions from the terms $\Lambda_{1}^{(0)}DT_{1}^{(1)}$ and $\Lambda_{1}^{(1)}DT_{1}^{(0)}$, and the fact that $\Lambda_{2}^{(0)}DT_{1}^{(1)}$ does not contribute in the RNCCSD method. Moreover, in the RCCSD method, the positive values of $T_{2}^{(0)}DT_{2}^{(1)}$ and $T_{2}^{(1)}DT_{2}^{(0)}$ partially balance out the other higher-order terms (labelled `Others') that are negative, leading to a low value of $\alpha_d$ for it, as compared to the RNCCSD method. In case of the RNCCSD method, the cancellations are not only lesser, but also contain higher-order terms that add to the net contribution. The table, thus, serves as an illustration of the rich and complex meshing of terms that eventually leads to the observed differences between the RCCSD and the RNCCSD values of $\alpha_d$. In other words, the extra correlation contribution from RNCC method to the final value is in fact a consequence of lesser cancellations. In Fig. \ref{fig1}, we demonstrate pictorially the trends of contributions from various RCCSD and RNCCSD terms to $\alpha_d$ for better understanding of these differences. 

\begin{figure}[t]
\centering
\includegraphics[width=8.5cm, height=8.5cm]{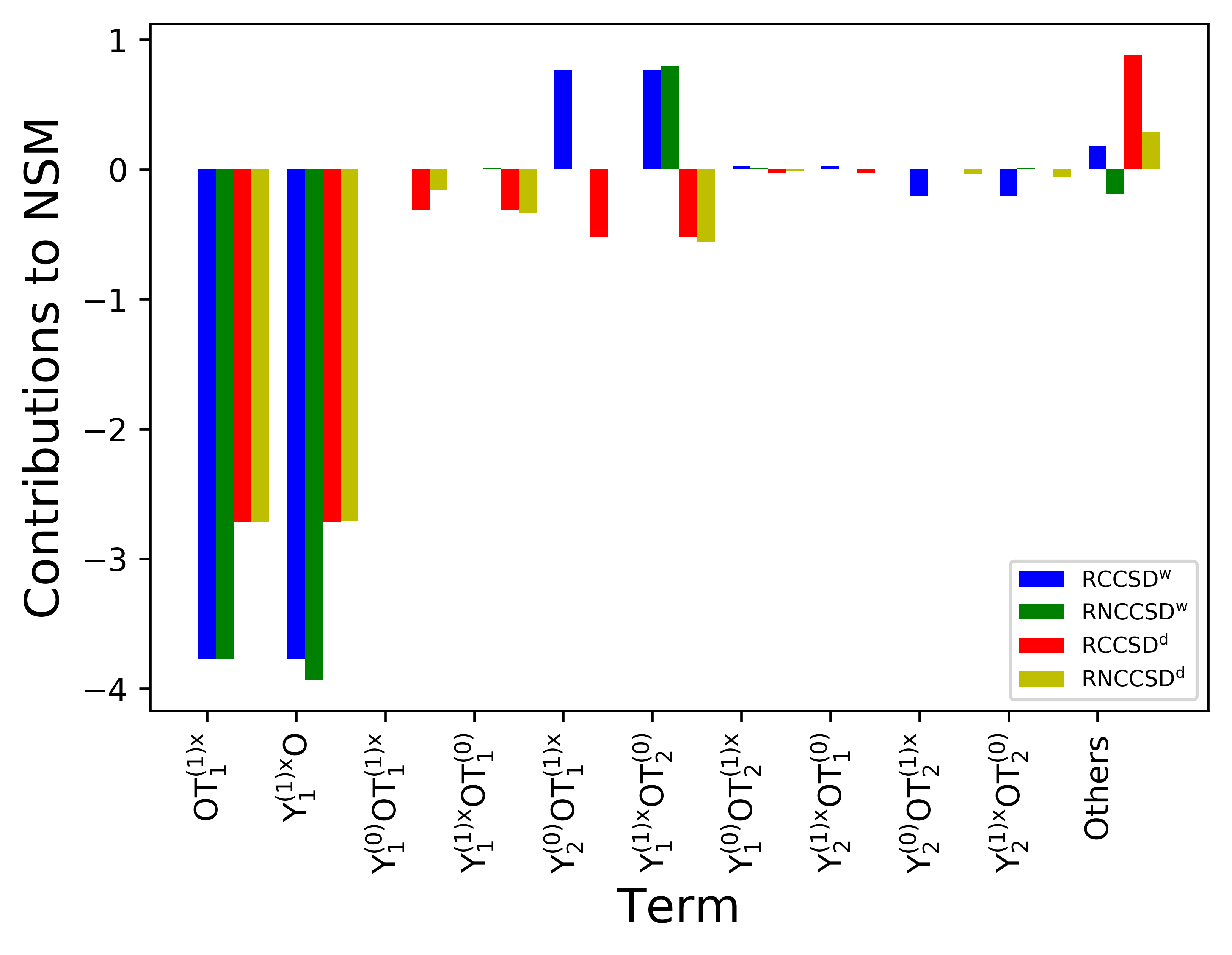}
\caption{A bar graph of contributions from various RCCSD and RNCCSD methods to $d_a$ due to NSM interaction in $^{225}$Ra. Values are scaled by ($10^{17} S \arrowvert e \arrowvert cm/(\arrowvert e \arrowvert fm^3)$). We adopt the same notations as in the previous figures here, with the operator $O$ being either the NSM interaction Hamiltonian or the dipole operator, depending on the choice for the perturbation. } 
\label{fig3}
\end{figure}

In a similar spirit, in Table \ref{tab:table3}, we shall now examine carefully the additions and cancellations at play among the individual terms that lead to the final $d_a$ value arising due to the T-PT e-N interaction Hamiltonian. The obtained results by perturbing with this P,T-odd interaction Hamiltonian and with the dipole operator $D$ are given side by side in the table, for ease of comparison. This can help in a comprehensive understanding of the roles of various correlation effects and to find out plausible reasons for differences in the results from both the approaches. For convenience, we shall refer to the R(N)CCSD results obtained by perturbing through weak interaction Hamiltonian as R(N)CCSD$^w$, and by R(N)CCSD$^d$ otherwise. By examining the table, one can see why results from the RCCSD$^w$ approach is the highest among the four approaches. We shall first look into the dominant contributions, which come from from $D {T_1^{(1)}}$, and its h.c. for the RCCSD method or $\Lambda_1^{(1)}D$ for the RNCCSD method, very much like in the case of $\alpha_d$. We find that although the dominant contribution ($D{T_1^{(1)w}}$, and its h.c.) in the RCCSD$^w$ approach is more negative ($-$12.934 units) than those ($D{T_1^{(1)d}}$, and its h.c. from RCCSD$^d$ and $D{T_1^{(1)d}}$ and $\Lambda_1^{(1)d}D$ from RNCCSD$^d$) from the two R(N)CCSD$^d$ methods ($-$10.854 units for RCCSD$^d$ and $-$11.181 for RNCCSD$^d$), this is offset by positive contributions from $T_{2}^{(0)}DT_{1}^{(1)w}$ and $T_{1}^{(1)w}DT_{2}^{(0)}$ to $-$9.622 units, while consistent negative contributions leave the RCC$^d$ and the RNCC$^d$ values at $-$12.798 and $-$12.214 units, respectively. We find that the rest of the contributions are negligible for our purposes (due to cancellations among themselves), and therefore conclude that the RCCSD$^w$ value is greater than its R(N)CCSD$^d$ counterparts, due to the cancellation between these two dominant contributions being very pronounced. Lastly, we shall comment on the RNCCSD$^w$ value, which is the lowest, at $-$13.148 units, by only invoking the two most dominant terms (the rest of the terms do not affect the analysis). We observe that the cause of a low net value, besides others discussed above, is due to two reasons: $\Lambda_1^{(1)w}D$ being more negative than its RCC counterpart, and also $\Lambda_2^{(0)\dagger}DT_1^{(1)w}$ being zero. These trends are also reflected in Fig. \ref{fig2} and leads to visualizing clearly the importance of correlation effects from various terms. Also, comparing this figure with Fig. \ref{fig1} can provide insights into the differences in the radial behavior of dipole operator and T-PT e-N interaction Hamiltonian. 

We now present individual contributions to $d_a$ due to the NSM interaction from the RCCSD and RNCCSD methods in Table \ref{tab:table4}. These are also depicted in Fig. \ref{fig3} for their qualitative behaviour, but as in the previous analysis, we need to take recourse to arithmetic to be able to explicitly identify the dominant terms systematically. For example, a cursory glance at this figure and comparing with Fig. \ref{fig2} indicates that the trends are almost similar for both the T-PT and NSM interactions in the evaluation of $d_a$ in $^{225}$Ra. However, there are still some notable differences that stand out, leading to all the four values being relatively close to each other and the results from R(N)CCSD$^d$ matching very well with each other as compared to their R(N)CCSD$^w$ counterparts. The first difference between the trends in the T-PT and NSM is that the $D{T_1^{(1)w}}$, and its h.c. from the RCCSD method are very close to  $D{T_1^{(1)w}}$ and $\Lambda_1^{(1)w}D$ from the RNCCSD method ($-$7.54 and $-$7.701 units, respectively), while the values are even closer for the R(N)CCSD$^d$ case ($-$5.44 and $-$5.425 units, respectively). Since the final values for $d_a$ due to the NSM obtained from RCCSD$^d$ and RNCCSD$^d$ are very close to each other, one would naively expect the second most dominant contributions to be almost equal. However, this is not true. One sees that although the sum of the rest of the terms except `Others' are very different ($-$1.724 and $-$1.159 units from RCCSD$^d$ and RNCCSD$^d$, respectively), and so are the contributions from `Others' (0.880 and 0.290 units from RCCSD$^d$ and RNCCSD$^d$, respectively), the combined sum in each of these methods are very similar, and remarkably, when they are combined further with the dominant contribution ($D{T_1^{(1)w}}$, and its h.c. from the RCCSD method and $D{T_1^{(1)w}}$ and $\Lambda_1^{(1)w}D$ from the RNCCSD method), the two final values almost match. On the other hand, terms except the most dominant ones do not add up symmetrically for R(N)CCSD$^w$ cases, and give 1.356 and 0.649 units for RCCSD$^w$ and RNCCSD$^w$, respectively, thereby causing the disagreement between their final values. 

\section{Conclusions} \label{sec5} 

In this work, we employ the relativistic normal coupled-cluster (RNCC) theory to verify the previous relativistic coupled-cluster (RCC) calculations on the electric dipole moment (EDM) of $^{225}$Ra due to parity and time-reversal violating nuclear Schiff moment and the  tensor-pseudotensor electron-nucleus interactions. The problems of unequal cancellation in the  normalization of the wave function and appearance of a non-terminating series in evaluating the EDM in the RCC theory are circumvented by the RNCC theory. We find large differences among the results for EDM between the RCC and RNCC theories at the level of singles and doubles excitations approximations. To verify the accuracies of our calculations, we have also determined dipole polarizabilities using both RCC and RNCC theories, and compared them with the results that have been obtained previously by employing other RCC methods. Further, we evaluated EDMs by formulating a mathematically equivalent expression for them by perturbing the wave function with the dipole operator instead of the P,T-odd interaction. Significant differences were observed between the results obtained using both the approaches. However, we notice good agreement in the EDM results that were evaluated by the RCC and RNCCSD theories when they are perturbed by the dipole operator. This suggests that less numerical uncertainties are introduced to the calculation of EDM in this procedure. This test is important to test the reliability in the calculations of EDMs in atomic systems. By analyzing the polarizability results and testing their reliability through the above procedure, we recommend enhancement factors to the EDM of $^{225}$Ra due to NSM and T-PT interactions as $-6.29(1) \times 10^{-17} \arrowvert e \arrowvert$cm $(\arrowvert e \arrowvert fm^3)$ and $-12.66(14) \times {10^{-20} \langle \sigma_N \rangle \arrowvert e \arrowvert }$cm, respectively. 

\section*{Acknowledgments}
Computations were performed on the VIKRAM-100 cluster at Physical Research Laboratory, Ahmedabad, India.

\end{document}